\documentclass{article}
\usepackage[margin=1in]{geometry}
\usepackage[utf8]{inputenc}
\usepackage{amsfonts}

\title{Tutorial on algebraic deletion correction codes}
\author{Kedar Tatwawadi\thanks{Corresponding author: \href{kedart@stanford.edu}{kedart@stanford.edu}}, Shubham Chandak\\Stanford University}
\date{April 2019}

\usepackage[numbers]{natbib}
\usepackage{graphicx}
\usepackage{amsmath}
\usepackage{hyperref}

\newtheorem{definition}{Definition}
\newtheorem{lemma}{Lemma}

\begin{document}

\maketitle

\begin{abstract}
    The deletion channel is known to be a notoriously difficult channel to design error-correction codes for. In spite of this difficulty, there are some beautiful code constructions which give some intuition about the channel and about what good deletion codes look like. In this tutorial we will take a look at some of them. This document is a transcript of my talk at the coding theory reading group on some interesting works on deletion channel. It is not intended to be an exhaustive survey of works on deletion channel, but more as a tutorial to some of the important and cute ideas in this area. For a comprehensive survey, we refer the reader to the cited sources and the following papers and surveys \cite{levenshtein1966binary,mitzenmacher2009survey, sloane2000single}.
    
    We also provide implementation of VT codes that correct single insertion/deletion errors for general alphabets at \url{https://github.com/shubhamchandak94/VT_codes/}.
\end{abstract}
\section{Introduction}
Deletion channel is one of the most fundamental of the channels, and is still not well understood. Most of you would know what a deletion channel/error looks like. But, to give an example, this is what a single deletion looks like: 

\begin{equation*}
    10110 \rightarrow 1010
\end{equation*}

Here the decoder receives the message $1010$ and needs to recover the original message. This corresponds to a single deletion, either in position 3 or 4. We cannot say what position the deletion occurred. 

In relation to our favourite erasure channel, where such a single error might look like:

\begin{equation*}
    10110 \rightarrow 10e10
\end{equation*}
 Here, we know what positions the erasure happened. The deletion channel is in fact a strictly worse channel, as we can convert the erasure channel output into a deletion channel output, by simply removing all the `$e$' symbols from the output.

We will also see that it has connections to our second favourite channel the Binary Symmetric Channel (BSC) channel. 

\subsection{Capacity of Deletion Channel}

We define a binary deletion channel, BDC with deletion rate $\alpha$, where each symbol in input $x^n$ can get deleted with probability $\alpha$. This definition has similarities with the BEC and BSC, but that's where the similarity ends. Surprisingly unlike BEC and BSC, we know quite less about BDC. 

\begin{enumerate}
    \item We still do not know the capacity of the BDC channel. One reason is that BDC is not really a discrete memoryless channel (DMC). For a DMC, you should be able to write:
    \begin{equation*}
        p(y^n|x^n) = \prod_{i=1}^n p(y_i|x_i)
    \end{equation*}
    We cannot write the same for deletion channels, one reason: as the output does not have ``length" $n$, and is in fact a random variable. Shannon theory gives us a nice characterization for the capacity of DMCs:
    
    \begin{equation}
        C = \max_{p(x)} I(X;Y)
    \end{equation}
    But, without this nice expression, finding the capacity is a much more difficult task. 
    \item We can of course get some bounds on the deletion channel capacity. For example: we know that binary erasure channel is strictly better than the BDC. Thus: 
    
    \begin{equation}
        C_{BDC}(\alpha) \leq 1 - \alpha
    \end{equation}
    
    This has been an open problem since quite some time now, but there has been some recent progress on this, which I will briefly talk about. 
    
    Most of the results are a part of the survey by Mitzmacher \cite{mitzenmacher2009survey}. Kalai, Mitzenmacher, and Sudan \cite{kalai2010tight} showed that when $\alpha \rightarrow 0$ the capacity is almost equal to the BSC. Best known capacity lower bounds I am aware of are \cite{drinea2007improved}:  $0.11(1-\alpha)$ . Recently, improved capacity upper bounds were obtained by Cheraghchi \cite{cheraghchi2019capacity} which state that:
    
    \begin{equation}
        C_{BDC} \leq (1-\alpha)\log \Phi 
    \end{equation}
    
    for $\alpha \geq 0.5$, where $\Phi$ is the golden ratio. For those of you who are interested, they essentially find capacity bounds for a channel known as Poisson repeat channel (the number of times each symbol is repeated is a Poisson random variable), which are then ported over for deletion channel as a special case. 
    
    This is mainly to give a sense of how difficult the deletion channel analysis is, and how little we really know about it. Typically, not knowing the capacity directly translates to not knowing good codes. But, surprisingly we know some nice code constructions in specific cases, which we will discuss next. 
    
    Deletion channel is also related to the insertion channel, or the indel channel, where both insertions and deletions happen, or general edit distance channels: indels + substitutions/reversals. A better understanding of the deletion channel is useful not just for communication, but for the problem of denoising, which is quite common (say when we type things and miss a character). 
    
\end{enumerate}

\section{Adversarial deletion error}
We spoke about ``Shannon-type" random deletion error capacity. But, for majority of the talk, we will mainly talk about the adversarial error, where when we say $e$ errors, we mean at max $e$ symbols are deleted. In this context let us define an $e$-deletion error correction code. 

\begin{definition} For any vector $u \in \mathrm{F}_2^n$ we define $D_e(u)$ as the descendants; i.e. all the vectors in $\mathbb{F}_2^{n-e}$ obtained after $e$ deletions.  
\end{definition}

\noindent For example: \\
For $u=01101$, $D_1(u) = \{ 0110, 0111, 0101, 1101\}$\\
For $u=00000$, $D_1(u) = \{ 0000\}$\\

Thus the descendant balls need not be of the same sizes unlike the Hamming balls we are familiar with. And this will lead to some interesting scenarios as we will see. In fact we can analyze the size of 1-deletion balls: 

\begin{lemma}
The size of 1-deletion Descendant ball $D_1(u^n)$ of a vector $u^n$ is equal to the number of ``runs" in $u^n$.  
\end{lemma}

\noindent For example: \\
$u^7 = 0010111$ has $4$ runs in total $00, 1, 0, 111$. Thus, $|D_1(u^7)| = 4$. It is easy to see why this is true: Any deletion in a run essentially leads to the same $n-1$ length sequence in the descendant ball. 

\begin{definition} We call a subset $C \subset \mathbb{F}_2^n$ to be a $e$-error correction code, if for any $u,v \in C, u \neq v$, $D_e(u) \cap D_e(v) = \phi$
\end{definition}

We will mainly deal with $e=1$ during the talk. 

\subsection{Repetition coding}
Lets start with the simplest of the $1$-deletion correction codes you can think of: ``repetition codes". We repeat every symbol $2$ times. Is that sufficient? Let us say, we observe some symbol an odd number of times, then we know that a symbol was deleted in that runs of $0$'s or $1$'s. Note that, we still cannot figure out the location of the deletion, but can figure out what was deleted.

This idea can in fact be extended to correct $e$ deletions by repeating every symbol $e+1$ times. But, this is quite bad: For correcting $1$ error, our communication rate is $0.5$. Still better that the BSC case, where we had to repeat things 3 times. 

Cool! Can we do better? We will next look at a cool class of codes known as the VT-codes; or the Varshamov-Tenengolts  code. But before doing let's look at a puzzle.

\subsection{ A puzzle}

I believe the puzzle has some connection with VT-codes, might help with the understanding; but if not, it is a cool simple puzzle! So lets say, Mary is the Queen of the seven kingdoms, and she had ordered 100 big barrels filled with gold coins, where each coin has a weight of 10gm. But, she knows from her secret agency that one of the barrels contains counterfeit coins weighing only 9gm. She has an electric weighing scale which she can use; so the question is: 

\emph{ How can she determine which barrel contains the counterfeit coins with a single measurement}

The solution is simple: she takes $j$ coins from the $j^{th}$ barrel and places them on the electronic weighing scale. Now if the weight is less that expected by $r$ grams, then it is the $r^{th}$ barrel which is counterfeit! We will come back to this puzzle :)

\section{VT-codes} 

Allright! We are all set to define VT-codes. 

\begin{definition}
Varshamov-Tenengolts code $VT_a(n) \subset \mathbb{F}^n_2$ is defined as:
\begin{equation}
    VT_a(n) = \left\{ x^n | \sum_{i=1}^n ix_i \equiv a \mod (n+1) \right\}
\end{equation}
\end{definition}

Some historical context on these codes: These codes were first proposed as error correction codes for 1-bit Z-channel error! Which means essentially $1\rightarrow 0$, but not the other way round. Z-channel errors are known as asymmetric bit-flips. Varshamov and Tenengolts proposed these codes in 1965 \cite{varshamov1965codes}, and then Levenshtein discovered that these codes in fact work well also for the deletion channel as well! 

\subsubsection*{Z-channel correction}
So before we look into 1 deletion correction, let us see how they can correct one Z-channel error, i.e. one of the $1$'s can flip to a $0$. 

Let $\hat{x}^n$ be the received symbol: then we can still compute:
\begin{equation*}
    S = a - \sum_{i=1}^n i\hat{x}_i 
\end{equation*}
In case, there is a $1 \rightarrow 0$ flip at position $j$, then $S=j$. Thus, we can correct the Z-channel error! Here is where the similarity with the puzzle can be seen.

\subsection{ VT-codes decoding}

We are all set to discuss the decoding for deletion channel: 

\begin{enumerate}
\item First of all, note that if it is only a deletion channel, then ``error detection" comes for free from the length of the code, unlike the bit-flipping error
\item Let $y^{n-1}$ be the received erroneous codeword after $1$ deletion. Define:

\begin{align}
    w &= \sum_{i=1}^{n-1} y_i \\
    S &= a - \sum_{i=1}^{n-1} iy_i
\end{align}

\item Let $p$ be the position at which deletion occurred, as in $x_p$ was deleted. Let $L_0, L_1$ be the counts of 1,0 to the left of position $p$. and $R_0, R_1$ be the counts to the right. 
In that case: $S = R_1$ if $x_p = 0$ and $S = p + R_1$ if $x_p = 1$. 

\begin{align*}
    S &= p + R_1 \\
      &= L_0 + L_1 + 1+ R_1 \\
      &= w + L_0 + 1 
\end{align*}

Thus, as $R_1 < w + 1$, if $S \leq w$, $0$ is deleted at a position with $R_1$ 1's to its right, otherwise, $1$ is deleted such that there are $L_0 = S -w -1$ zeros to its left. 
\item Note that we can thus uniquely determine the $x^n$ sequence, but we cannot determine the exact location of the deletion here, as it can be any $0$ or $1$ in the run we identified.

\end{enumerate}

This beautiful decoding algorithm was given by Levenshtein in his 1965 work \cite{levenshtein1966binary}. Also most of the things which I will talk are from the survey by Sloane \cite{sloane2000single}. 

\subsection{VT-codes rate}
As the VT-codes are combinatorial, we can get exact formulae for their sizes. Exact combinatorial formulate can be found in \cite{sloane2000single}. Here are few interesting things: 

\begin{enumerate}
    \item 
    \begin{lemma}
    For some $a \in [0,n]$:
    \begin{equation*}
        |VT_a(n)| \geq \frac{2^n}{n+1}
    \end{equation*}
    \end{lemma}
    As every $x^n \in \mathbb{F}_1^n$ lies in exactly one of $VT_a(n)$ sets:
    
    \begin{equation*}
        \sum_{a=0}^n |VT_a(n)| = 2^n
    \end{equation*}
    This leads to the lemma. 
    \item It can be in fact shown that $a=0$ leads to the largest code size, and $a=1$ the smallest size. 
    \begin{equation*}
        |VT_0(n)| \leq |VT_a(n)| \leq |VT_1(a)|
    \end{equation*}
    \item For $n=2^q-1$, all the VT-code sizes are in fact equal to $\frac{2^n}{n+1}$. 
\end{enumerate}

\subsection{Optimality of VT-codes}
We say a $e$-error correction code is ``optimal" if it has the smallest size amongst all $e$-error correction codes. Let us analyze the ``optimality" of VT-codes. 
\begin{enumerate}
    \item Levenshtein \cite{levenshtein1966binary} showed that, optimal $1$-deletion correction codes have asymptotic sizes  $\approx \frac{2^n}{n}$. This makes \textit{VT-codes} asymptotically optimal. 
    \item People have not been able to prove that VT-codes are optimal non-asymptotically. Finding the ``optimal" 1-deletion code is a NP-hard problem, as it involves finding the independent set on the graph where vertices are connected if they lead to the same deletion descendants \cite{sloane2000single}. But for $n \leq 8$ it is known that they are optimal, using computer programs. 
    Sizes of these codes are:
    \begin{equation*}
        1,2,2,4,6,10,16,30
    \end{equation*}
     For higher $n$ due to the exponential nature of the algorithms, we cannot say anything yet. 
    \item VT-codes also have the property that they are ``perfect codes" \cite{levenshtein1992perfect} which implies that their descendent sets which are disjoint cover the entire $2^{n-1}$ sized. For example:
    
    \begin{align*}
        VT_0(3) &= \{000,101 \} \\
        Descendants &= \{ 00, 10, 01, 11 \}
    \end{align*}
    
    Levenshtein showed that surprisingly, this is true for all $a$, which is quite cool in itself! The perfect codes analogy comes from Hamming codes being Perfect. But, unlike the Hamming distortion case, here perfect codes does not imply optimal codes? Why? For example, code $\{000,111\}$ is not perfect but is perhaps a better code that $VT_0(3)$, potentially there might be a larger code for larger $n$. Why does this happen? Because, the number of descendants are not fixed, some have $1$ and some have more. 
    
    \item \textbf{Linearity}: VT codes are linear until $n=4$ but never after! \cite{sloane2000single}. Variants of VT-code (restrictions on VT-codes) can be made linear by considering redundancy to be $\sqrt{n}$ as against $\log n$. Althought I am not sure how is linearity of codes useful, if the decoding is still non-linear (linear time complexity, but non-linear in nature). 
\end{enumerate}

\subsection{Systematic Encoding}
Now that we have taken a look at the linear-time decoding of VT-codes, it is a natural question to ask if there exist a nice way to encode data. This problem surprisingly remained open for more than 30 years until 1998 when Abdel-Gaffar et al. \cite{abdel1998systematic} provided a very convenient way of in fact ``systematic encoding" of data. 

\begin{enumerate}
    \item For $n = 2^q - 1$, let $m = 2^q - q - 1$ be the number of data bits, and $q$ are the ``parity" bits.
    Let the data bits be $b_1,b_2,\ldots, b_m$, and the code-word to be formed $x^n$ in $VT_a(n)$.
    \item Fill in the data except in positions $1,2,2^2,2^3,\ldots, 2^{q-1}$. Thus $x^n$ codeword looks like: 
    \begin{equation*}
        x_1, x_2, b_1, x_4, b_2, b_3, b_4, x_8, \ldots, b_m
    \end{equation*}
    
    We can compute: $S = a - \sum_{i=1}^{2^q - 1} ix_i \mod (2^q)$. As most of the positions of $x^n$ (except $x_{2^j}, j \in [0,q-1]$) are decided, to obtain $S = 0$, we need:
    \begin{equation*}
        \sum_{j=0}^{q-1} 2^j x_{2^j} \mod (2^q) = \hat{a}
    \end{equation*}
    We can now conveniently choose $x_{2^j}, j \in [0,q-1]$ as the $q$-bit binary expansion of $\hat{a}$.
    \item Note that $m = 2^q -q -1, n = 2^q-1$ for 1-deletion correction is exactly same as the rate of Hamming code. Not sure if this is a coincidence or something more!
\end{enumerate}

\section{Insertion + Deletion + Substitution codes}
We looked at $1$-deletion correction channels in depth. In the next part of the tutorial, we will extend this understanding to more general scenarios. The first scenario is $1$ insertion instead of $1$ deletion error. 
\subsection{General indel error codes}
Levenshtein \cite{levenshtein1966binary} showed this general lemma: 
\begin{lemma}
Any $s$-deletion correction code can also correct $s_1$ deletions and $s_2$ insertion errors where $s_1 + s_2 \leq s$.
\end{lemma}
Note that here we do not need to know $s_1,s_2$ beforehand. The general proof is quite simple. Here, we will prove the simpler version of $1$-insertion error, as that is sufficient to get an intuitive understanding. 

\begin{enumerate}
\item Let us assume that $C$ is a $1$-deletion correction code. We want to show that $C$ can correct $1$-insertion errors as well. 
\item Let us assume that on the contrary, there exist codewords $x^n, y^n \in C$ such that after one insertion error in them the resulting noisy codewords $\hat{x}^{n+1},\hat{y}^{n+1}$ are equal. Let the insertion occured at position $i$ in $\hat{x}^{n+1}$ and at position $j$ in $\hat{y}^{n+1}$. 
\item As $\hat{x}^{n+1} = \hat{y}^{n+1}$, even after deleting symbols in position $i,j$ in both $\hat{x}, \hat{y}$, results in $n-1$ vectors which are equal. However, the $n-1$ length codewords are in fact $1$-deletion descendants of the codewords $x^n, y^n \in C$. This is contradictory to the definition of deletion correction codes, as no descendants can be equal. Thus, $C$ has to be $1$-insertion correction as well.
\end{enumerate}

Note that, this is more of an existential result, and efficient deletion error decoding algorithms might not translate into efficient insertion detection algorithms.

\subsection{VT-codes for 1-insertion, deletion, substitution correction}
Levenshtein showed the surprising fact that with a simple modification standard VT-codes can be converted into 1-insertion, deletion and substitution correction codes. The modification is as follows:

 $\hat{VT}_a(n) \subset \mathbb{F}^n_2$ is defined as:
\begin{equation}
    \hat{VT}_a(n) = \left\{ x^n | \sum_{i=1}^n ix_i \equiv a \mod (2n+1) \right\}
\end{equation}

Let us try to understand why $\hat{VT}_a(n)$ work:
\begin{enumerate}
    \item First of all, from the length of the code, we know whether there is an insertion, deletion or a substitution.
    \item Recollect that deletion error correction in $VT_a(n)$ codes only depends on distinct remainders modulus $(n+1)$. This should still hold true if the modulus is taken $m \geq n+1$. Thus, with $m=2n+1$, 1-deletion correction still holds. 
    \item 1-insertion code ability was already shown from the general lemma earlier. However, using a similar remainder trick $\mod 2n+1$, insertions can in fact be corrected efficiently using the $\hat{VT}_a(n)$ codes.
    \item The only case remaining to analyze is the 1-substitution or 1-bitflip case. Let $0\rightarrow 1$ at position $p \in [1,n]$ in the codeword $x^n$ resulting in the noisy codework $\hat{x}^n$. Clearly: 
    \begin{align*}
        S &= a - \sum_{i=1}^n i\hat{x}_i \mod (2n+1) \\
          &= 2n-p\\
    \end{align*}
    
    If $1 \rightarrow 0$, then: 
    \begin{align*}
        S &= a - \sum_{i=1}^n i\hat{x}_i \mod (2n+1) \\
          &= p\\
    \end{align*}
    As all these $S$ values are distinct, we can correct for 1-bitflip. Note that, this construction is not optimal just for 1-bitflip, as it essentially encodes 1 bit less than Hamming codes.
\end{enumerate}



\section{VT-codes for larger alphabet}
Creating deletion codes for larger alphabets becomes a bit tricky. Of course, repetition coding still works on non-binary alphabets. 
\subsubsection*{Code which does not work}

When I started thinking about this problem, I came up with this code, which has a bug! Let us still take a look at it, as it gives some understanding as to the intricacies of code-design:  

Define $\beta_i$ for $i \in [2,n]$ as:
\begin{align*}
    \beta_i  &= 1,\ x_i \neq 0 \\
              &= 0,\  \mathrm{otherwise} 
\end{align*}

Then we consider code as:

\begin{align*}
    C = \bigg {\{}  x^n | &\sum_{i=1}^n i\beta_i = a \mod (n+1), \\
      & \sum_{j=1}^n x_i = b \mod (q)\bigg{\}}
\end{align*}

Let us look at the argument as to why the code works, and try to find the bug! \\
\emph{\textbf{Argument:} The first equation, similar to the binary VT code will tell us the position of the deletion, and the second equation tells us the value. }

Why does the above argument not work. The reason is that binary VT-codes not not actually tell us the position of the deletion correctly. They can tell us in which ``run" the deletion happened, and hence obtain the codeword correctly, but not the exact position. 

How do we solve for that?
\subsubsection*{Code which works}
This code appears in the work of Tenengolts in 1984. \cite{tenengolts1984nonbinary}\\
Define $\alpha_i$ for $i \in [2,n]$ as:
\begin{align*}
    \alpha_i  &= 1,\ x_i \geq x_{i-1} \\
              &= 0,\  x_i < x_{i-1} 
\end{align*}

Then we consider the code:

\begin{align*}
    C = \bigg \{  x^n | &\sum_{i=2}^n (i-1)\alpha_i = a \mod (n) , \\ 
      &\sum_{j=1}^n x_i = b \mod (q) \bigg \}
\end{align*}

Let us try to analyze the decoding for this code: 

\begin{enumerate}
    \item First of all, as in the previous code, from the second constraint $\sum_{j=1}^n x_i = b \mod (q)$, we figure out what is the value of the deleted symbol. What remains to determine is its position.
    \item The $\alpha$ sequence is essentially capturing the monotonous regions of the $x^n$ sequence. $\alpha_i = 1 $, when the sequence is increasing (non-decreasing), and $0$ when it is decreasing. Thus, a deletion in $x^n$ sequence will in fact lead to exactly $1$ deletion in the $\alpha$ sequence (the position of deletion might be shifted by 1, but it does not matter as VT-codes do not correct for position). 
    \item As the first equation is a VT-code, it can determine the run in which the deletion occurred. As every run in $\alpha$ sequence corresponds to monotonic increasing/decreasing subsequence in $x^n$, from the value of the deleted symbol, we can correctly place it and complete the decoding!
\end{enumerate}

Efficient systematic encoding for this code was discovered recently in a work by Abroshan et al.\ \cite{abroshan2018efficient} and is included in the implementation at \url{https://github.com/shubhamchandak94/VT_codes/}.
\section{Bursty deletion codes}

In this section we will look at Bursty deletions. By a single bursty deletion of size $s$, we mean that some consecutive $s$ symbols were deleted. Note that, $s$-bursty deletion correction codes can correct for exactly 1 burst of size $s$, but surprisingly they need not correct for a burst of size $s-1$. For example: $C = \{0101, 1010 \}$ can correct 2-bursty errors, but not single deletion errors!

\subsection{VT-code based construction}

We consider a construction based on single-deletion correction VT-codes to correct a single burst of  $s$ deletions. How should one do that? One simple trick is to distribute these $s$ deletions across the $n$ length sequence, so that each $n/s$ length subsequence has exactly 1 deletion. For simplicity, let $n/s = k$.

Then the codeword $x^n$ has the property that each of the $s$ rows below, belong to a $VT_a(n)$ code. 
\begin{align*}
    &x_1, x_{s+1}, x_{2s+1}, \ldots, x_{n-s+1} \\
    &x_2, x_{s+2}, x_{2s+2}, \ldots, x_{n-s+2} \\
    &\;\;\;\;\vdots \\
    &x_s, x_{2s}, x_{2s+2}, \ldots, x_{n} \\
\end{align*}

\begin{enumerate}
    \item Let us analyze the scenario of 1 bursty error of size $s$ from position $x_j, \ldots, x_{j+s-1}$ are deleted. This corresponds to exactly 1 deletion in each of the rows.
    \item We still need to figure out which symbols of the deleted codeword belong to which rows, as the alignment might no longer be true. The cool thing is that, if there are \emph{exactly} $s$ consecutive deletions, then every $s^{th}$ symbol will still be correctly aligned to the rows. 
    \item Thus, our code can in fact correct a bursty error of $s$ deletions, but need not correct lower number of bursts, which is quite unusual!
\end{enumerate}  

One important caveat to observe here is that the position of deletion in each of the rows is the same, or shifted by 1. Thus, if one of the rows is a VT-code, and the other rows just tell whether the error position is odd or even, that is enough to resolve the bursty error. This observation is the basis of further improvements to bursty error correction. For more details take a look at this paper: \cite{schoeny2017codes}.
\section{Multiple Deletions}

One would imagine that it should be possible to extend the elegant construction of VT-codes from single deletion correction to multiple deletions. However, this problem has proved to be much more difficult. 

There have been some recent works which extend the single deletion errors to multiple errors. Gabrys et al \cite{gabrys2018codes} provide an extension of the VT-codes idea to correct two deletions. There have been other recent works which provide multiple deletion correcting codes using different (non VT-code based) ideas \cite{brakensiek2017efficient}. 



\section*{Acknowledgement}
I would like to thank Jay Mardia and Mary Wootters for interesting discussions on deletion codes. 

\begin{thebibliography}{10}

\bibitem{abdel1998systematic}
Khaled~AS Abdel-Ghaffar and Hendrik~C Ferreira.
\newblock {Systematic encoding of the Varshamov-Tenengol'ts codes and the
  Constantin-Rao codes}.
\newblock {\em IEEE Transactions on Information Theory}, 44(1):340--345, 1998.

\bibitem{abroshan2018efficient}
Mahed Abroshan, Ramji Venkataramanan, and Albert Guillen~I Fabregas.
\newblock Efficient systematic encoding of non-binary vt codes.
\newblock In {\em 2018 IEEE International Symposium on Information Theory
  (ISIT)}, pages 91--95. IEEE, 2018.

\bibitem{brakensiek2017efficient}
Joshua Brakensiek, Venkatesan Guruswami, and Samuel Zbarsky.
\newblock Efficient low-redundancy codes for correcting multiple deletions.
\newblock {\em IEEE Transactions on Information Theory}, 64(5):3403--3410,
  2017.

\bibitem{cheraghchi2019capacity}
Mahdi Cheraghchi.
\newblock Capacity upper bounds for deletion-type channels.
\newblock {\em Journal of the ACM (JACM)}, 66(2):9, 2019.

\bibitem{drinea2007improved}
Eleni Drinea and Michael Mitzenmacher.
\newblock Improved lower bounds for the capacity of iid deletion and
  duplication channels.
\newblock {\em IEEE Transactions on Information Theory}, 53(8):2693--2714,
  2007.

\bibitem{gabrys2018codes}
Ryan Gabrys and Frederic Sala.
\newblock Codes correcting two deletions.
\newblock {\em IEEE Transactions on Information Theory}, 65(2):965--974, 2018.

\bibitem{kalai2010tight}
Adam Kalai, Michael Mitzenmacher, and Madhu Sudan.
\newblock Tight asymptotic bounds for the deletion channel with small deletion
  probabilities.
\newblock In {\em 2010 IEEE International Symposium on Information Theory},
  pages 997--1001. IEEE, 2010.

\bibitem{levenshtein1966binary}
Vladimir~I Levenshtein.
\newblock Binary codes capable of correcting deletions, insertions, and
  reversals.
\newblock In {\em Soviet physics doklady}, pages 707--710, 1966.

\bibitem{levenshtein1992perfect}
Vladimir~I Levenshtein.
\newblock On perfect codes in deletion and insertion metric.
\newblock {\em Discrete Mathematics and Applications}, 2(3):241--258, 1992.

\bibitem{mitzenmacher2009survey}
Michael Mitzenmacher et~al.
\newblock A survey of results for deletion channels and related synchronization
  channels.
\newblock {\em Probability Surveys}, 6:1--33, 2009.

\bibitem{schoeny2017codes}
Clayton Schoeny, Antonia Wachter-Zeh, Ryan Gabrys, and Eitan Yaakobi.
\newblock Codes correcting a burst of deletions or insertions.
\newblock {\em IEEE Transactions on Information Theory}, 63(4):1971--1985,
  2017.

\bibitem{sloane2000single}
Neil~JA Sloane.
\newblock On single-deletion-correcting codes.
\newblock {\em Codes and designs}, 10:273--291, 2000.

\bibitem{tenengolts1984nonbinary}
Grigory Tenengolts.
\newblock {Nonbinary codes, correcting single deletion or insertion
  (Corresp.)}.
\newblock {\em IEEE Transactions on Information Theory}, 30(5):766--769, 1984.

\bibitem{varshamov1965codes}
RR~Varshamov and GM~Tenengolts.
\newblock {Codes which correct single asymmetric errors (in Russian)}.
\newblock {\em Automatika i Telemkhanika}, 161(3):288--292, 1965.

\end{thebibliography}

\end{document}